\documentclass[aps,prd,superscriptaddress,12pt]{revtex4-2}
\usepackage{inputenc}
\usepackage{amsmath}
\usepackage{amsfonts}
\usepackage{amssymb}
\usepackage{graphicx}
\usepackage{natbib}
\usepackage{hyperref}
\usepackage{setspace}

\usepackage{float}
\usepackage{orcidlink}
\newcommand{\be}{\begin{equation}}
\newcommand{\ee}{\end{equation}}
\newcommand{\bea}{\begin{eqnarray}}
\newcommand{\eea}{\end{eqnarray}}

\begin{document}

\title{Exploring the Evolution of Nonlinear Electrodynamics in the Universe: A Dynamical Systems Approach}

\author{Ricardo García-Salcedo\orcidlink{0000-0003-0173-5466}}
\affiliation{Centro de Investigación en Ciencia Aplicada y Tecnología Avanzada, Unidad Legaria del Instituto Politécnico Nacional, Legaria 694, 11500, Ciudad de México, México.}
\author{Isidro Gómez-Vargas\orcidlink{0000-0002-6473-018X}}
\affiliation{Department of Astronomy of the University of Geneva, 51 Chemin Pegasi, 1290 Versoix, Switzerland.}
\author{Tame González\orcidlink{0000-0003-3234-1224}}
\affiliation{Dpto. Ingeniería Civil, División de Ingeniería, Universidad de Guanajuato, Gto., México.}
\author{Vicent Martinez-Badenes\orcidlink{0000-0002-4056-9627}}
\affiliation{Universidad Internacional de Valencia - VIU, 46002, Valencia, Spain.}
\author{Israel Quiros\orcidlink{0000-0002-0120-0624}}
\affiliation{Dpto. Ingeniería Civil, División de Ingeniería, Universidad de Guanajuato, Gto., México.}

\date{\today}

\begin{abstract}
This paper investigates the dynamics of cosmological models incorporating nonlinear electrodynamics (NLED), focusing on their stability and causality. We explore two specific NLED models: the Power-Law and the Rational Lagrangian. By employing a combination of dynamical systems theory and Bayesian inference, we aim to understand the viability of these kinds of models in describing the evolution of the universe. We present the theoretical framework of NLED coupled with general relativity, followed by an analysis of the stability and causality of various NLED Lagrangians. We then perform a detailed dynamical analysis to identify the ranges where these models are stable and causal. Our results show that the Power-Law Lagrangian model transitions through various cosmological phases from a Maxwell radiation-dominated state and evolving to a matter-dominated state. For the Rational Lagrangian model, including the Maxwell term, stable and causal behavior is observed within specific parameter ranges, with critical points indicating the evolutionary pathways of the universe. To validate our theoretical findings, we perform Bayesian parameter estimation using a comprehensive set of observational data, including cosmic chronometers, Baryon Acoustic Oscillation (BAO) measurements, and Supernovae Type Ia (SNeIa). The estimated parameters for both models align with the expected values for the current universe, particularly the matter density $\Omega_m$ and the Hubble parameter $h$. However, the parameters $\alpha$ and $b$ are not tightly constrained within the prior ranges. Our model comparison strongly favors the $\Lambda$CDM model over the NLED models for late-universe observations since the NLED model does not exhibit a cosmological constant behavior. Our results highlight the need for further refinement and exploration of NLED-based cosmological models to fully integrate them into the standard cosmological
framework.

\end{abstract}

\maketitle


\section{Introduction}

The present paradigm of cosmology includes an early period of inflationary expansion \cite{guth1981inflationary}, a stage of decelerated matter dominance \cite{peebles1993principles}, and accelerated expansion at late times \cite{riess1998observational,perlmutter1999measurements}. These distinct phases are well-supported by observational evidence and are essential components of the standard cosmological model \cite{weinberg2008cosmology}. These dynamics can be achieved by introducing various material contents into Einstein's theory of general relativity \cite{kolb2018early} or by appropriately modifying the theory itself \cite{clifton2012modified}.

Nonlinear electrodynamics (NLED) has been a subject of interest since the early 20th century, primarily due to its potential to address issues not explained by classical Maxwellian electrodynamics. One of the first significant models was introduced by Born and Infeld in 1934, which aimed to eliminate the singularities associated with the self-energy of point charges \cite{born1934foundations}. This model modifies the Maxwell Lagrangian to produce finite energy solutions for point charges, thus resolving the infinite self-energy problem. NLED theories have since been studied extensively for their ability to describe high-intensity electromagnetic fields and their applications in various areas of physics, including cosmology and astrophysics \cite{plebanski1970lectures, boillat1970nonlinear, novello2000singularities}. In cosmology, NLED can provide alternative mechanisms for inflation, structure formation, and the late-time accelerated expansion of the universe \cite{garcia2000born, novello2004nonlinear}. These models are particularly valuable for their potential to offer insights into the early universe's dynamics and the nature of dark energy and dark matter.

In the context of cosmological models, NLED has been extensively studied, offering alternative mechanisms for early universe inflation, structure formation, and late-time accelerated expansion. Various Lagrangians have been proposed to capture the nonlinear effects of electrodynamics in different cosmological scenarios. For instance, the Born-Infeld type Lagrangian, $L=-\beta^2 (\sqrt{1+\frac{F}{2\beta^2}-\frac{G^2}{16 \beta^4}}-1)$, has been analyzed in early universe models, FRW cosmologies, and Bianchi universes, with a focus on the squared sound speed $c_s^2$ \cite{garcia2000born, dyadichev2002non, moniz2002quintessence, elizalde2003born, dyadichev2005chaos, de2009classification, garcia2014comment, garcia2014no}. Another model, represented by $L=-\beta^2 [(1+\frac{\beta F}{\sigma}-\frac{\beta \gamma G^2}{2 \sigma})^{\sigma}-1]$, explores nonsingular universes and bouncing scenarios, demonstrating that the NLED behaves as a cosmological constant at early and late times \cite{novello2012extended, kruglov2017notes}. 

Additionally, the Lagrangian $L=-\frac{1}{\mu_o 4}F + \alpha F^2 + \beta G^2$ has been studied extensively in the context of magnetic universes, inflation, and singularity-free scenarios within FRW cosmologies \cite{de2002nonlinear, camara2004nonsingular, novello2005cosmological, sharif2017stability}. The Lagrangian $L=-\frac{1}{\mu_o 4}F + \alpha F^2$ has been analyzed primarily for early-time expansion, inflation, and singularity-free models in FRW cosmologies \cite{de2002nonlinear, camara2004nonsingular, camara2007nonlinearity, campanelli2008inflation, vollick2008homogeneous, de2009classification, garcia2014no}. 

Other notable Lagrangians include $L=-\frac{F}{2 \beta F +1}$, which has been studied in FRW and Bianchi I universes, showing early-time acceleration without singularities \cite{kruglov2015universe, otalora2018inflation, singh2018accelerating, kruglov2017inflation}, and $L=-F e^{-\alpha F}$, which addresses inflation, singularity-free scenarios, and the squared sound speed $c_s^2$ \cite{campanelli2008inflation, kruglov2016nonlinear}. These studies highlight the diverse applications and significant potential of NLED in addressing key questions in cosmology, from the early universe dynamics to the late-time accelerated expansion.

The stability and causality analysis in NLED models ensures their physical viability, especially when considering cosmological applications. One of the critical indicators of stability is the behavior of the squared sound speed (SSS), defined as $c_s^2 = dp/d\rho$ \cite{peebles2003cosmological, ellis2007causality}. A stable cosmological model requires a positive SSS ($c_s^2 > 0$) to prevent the uncontrolled growth of energy density perturbations, which leads to classical instabilities known as gradient instabilities \cite{yang2012dynamics}. Additionally, the SSS must not exceed the local speed of light ($c_s^2 \leq 1$) to avoid superluminal propagation, which would violate causality \cite{hawking1973large, wald2010general, adams2006causality}. Early works on NLED, such as \cite{novello2005cosmological}, laid the foundation for these analyses, and subsequent studies, including \cite{garcia2014no}, have provided detailed discussions on the bounds of the SSS and their implications for the stability and causality of NLED models. Ensuring these conditions are met is important for the theoretical consistency and observational compatibility of NLED-based cosmological models.

For instance, the Born-Infeld type Lagrangian, $L=-\beta^2 (\sqrt{1+\frac{F}{2\beta^2}-\frac{G^2}{16 \beta^4}}-1)$, has been extensively analyzed for its stability and causality properties in early universe models and FRW cosmologies, demonstrating its robustness under certain conditions \cite{garcia2000born, dyadichev2002non, moniz2002quintessence, elizalde2003born, dyadichev2005chaos, de2009classification, garcia2014comment, garcia2014no}. Similarly, the Lagrangian $L=-F e^{-\alpha F}$ has been studied for its implications in inflationary scenarios and singularity-free models, with specific focus on ensuring that $c_s^2$ remains within the acceptable range to avoid instabilities \cite{campanelli2008inflation, kruglov2016nonlinear}. Ensuring the stability and causality of NLED models through $c_s^2$ analysis is a theoretical exercise and a necessary step in validating these models against observational data and ensuring their applicability in describing the universe's evolution.

This study explores electrodynamics beyond Maxwell's to capture the full range of cosmic evolution. Specifically, we investigate NLED models as potential candidates to replicate the observed dynamics of the Universe. However, it is essential to note that not all NLED models are suitable for this purpose. A rigorous examination of their stability and causality properties is necessary to ensure their viability.

Our approach involves a comprehensive analysis combining the dynamical behavior of a homogeneous and isotropic cosmological model coupled with nonlinear electrodynamic radiation and a Bayesian analysis of the resulting cosmological parameters. The parameters of these models were fine-tuned using observational data from Type Ia supernovae, which serve as standard candles for measuring cosmic distances and expansion rates.

Despite thorough parameter fitting, our results indicate that none of the NLED models under consideration can adequately describe the current stage of accelerated expansion of the Universe. This finding underscores the challenges in developing NLED models that align with all phases of the cosmic timeline and highlights the need for continued exploration and refinement of theoretical models in cosmology.

This paper is organized as follows: Section II presents the theoretical framework of NLED coupled to general relativity. Section III analyzes the stability and causality of various NLED Lagrangians. Section IV performs a dynamical analysis of models with stable and causal Lagrangian densities. Section V provides Bayesian parameter estimation using observational data. Finally, Section VI summarizes the conclusions and suggests directions for future research.

\section{Nonlinear Electrodynamics Coupled to General Relativity}\label{sec-defi}

\subsection{Modified Friedmann Equations}

This paper will consider the Einstein gravitational equations coupled with nonlinear radiation, represented by a nonlinear Lagrangian. In this context, the four-dimensional (4D) action of gravity coupled to nonlinear electrodynamics is given by ($\frac{8\pi G}{c^4}=1$, geometrical units):
\be 
    S=\int d^4x\sqrt{-g}\left[ R+L_m+L(F,G) \right],
    \label{action}
\ee 
where $R$ is the curvature scalar, $L_m$ is the Lagrangian density of the background perfect fluid, and $L(F,G)$ is the gauge-invariant electromagnetic (EM) Lagrangian density, which is a function of the electromagnetic invariants:
\be
    F=F_{\mu \nu}F^{\mu \nu}=2(B^2-E^2), \;\;\;\;\;\;\; G=\frac{1}{2}\epsilon_{\alpha \beta \mu \nu}F^{\alpha \beta} F^{\mu \nu}=-4{\bf E}\cdot{\bf B}, \label{InvFG}
\ee 
where {\bf B}, {\bf E} being the magnetic induction and electric fields correspondingly and $F_{\mu \nu}=\partial_\mu A_\nu-\partial_\nu A_\mu$ with the electromagnetic potential $A_\mu$.

In electrodynamics, the Maxwell-Lagrangian density can be formulated differently, affecting the resulting equations of motion and the interpretation of physical quantities. Specifically, the Lagrangian density for the electromagnetic field can be written as $L = -F$ where $F=\frac{1}{4}F_{\mu \nu}F^{\mu \nu}$ or $L = -\frac{1}{4}F$ where $F = F_{\mu \nu}F^{\mu \nu}$. The first formulation, $L = -F$, includes the normalization factor within the definition of the invariant $F$, simplifying some mathematical expressions. However, this approach can complicate the interpretation of energy and momentum densities. On the other hand, the formulation $L = -\frac{1}{4}F$ explicitly includes the normalization factor in the Lagrangian, which aligns more closely with the standard conventions in classical electrodynamics, making the interpretation of physical quantities more straightforward. 

Given these considerations, we will adopt the formulation $L = -\frac{1}{4}F$ in this work. This choice facilitates a clearer comparison with standard electrodynamics and ensures our results are consistent with established physical interpretations. This formulation will be particularly useful in deriving and analyzing the equations of motion and the stability conditions for the nonlinear electrodynamics (NLED) models considered in this study.

The Born-Infeld theory was the first nonlinear extension of electromagnetism to avoid the singularities associated with infinite electric fields in Maxwell's theory \cite{born1934foundations}. The Born-Infeld Lagrangian can be expressed as:
\be
    L = b^2 \left( 1 - \sqrt{1 - \frac{1}{2b^2} F} \right),
\ee
where $F = F_{\mu \nu}F^{\mu \nu}$. Core concepts of NLED \cite{plebanski1970lectures, gaete2014remarks, gaete2017note}, their applications across various physics disciplines \cite{sorokin2022introductory}, and potential avenues for further research are explored. However, this work focuses solely on nonlinear electrodynamics as the material content for a cosmological model, excluding further research directions.

The corresponding gravitational field equations can be derived from the action (\ref{action}) by performing variations with respect to the spacetime metric $g_{\mu\nu}$ to obtain: 
\be
    G_{\mu\nu}=T_{\mu\nu}^m+T_{\mu\nu}^{EM},
\ee 
where $G_{\mu\nu}$ is the Einstein tensor containing all the geometric information, and
\bea
    T_{\mu\nu}^m&=&\left(\rho_m+p_m\right) u_{\mu}u_{\nu}-p_m g_{\mu\nu}, \nonumber \\ T_{\mu\nu}^{EM}&=&g_{\mu\nu}\,\left[L(F, G)-G L_{,G}\right]-4F_{\mu\alpha}F_\nu^{\;\;\alpha}\,L_{,F},
    \label{em_t}
\eea 
are the energy-momentum tensors for ordinary matter ($m$) and the nonlinear electromagnetic field ($EM$), respectively. Here, $\rho_m=\rho_m(t)$ and $p_m=p_m(t)$ are the energy density and barotropic pressure of the background fluid, respectively, and $u_{\mu}$ is the normalized ($u_{\mu}u^{\mu} = 1$) velocity of the reference frame where the fields are measured. Also, $L_{,X}\equiv dL/dX$ and $L_{,XX} \equiv d^2L/dX^2$, etc.

The variation of action (\ref{action}) with respect to the components of the electromagnetic potential $A_\mu$ results in the electromagnetic field equations, referred to as modified Maxwell equations:
\be
    \left(F^{\mu\nu},L_{,F}+\frac{1}{2}\epsilon^{\alpha\beta\mu\nu}F_{\alpha\beta}L_{,G}\right)_{;\mu}=0,
    \label{em-field}
\ee
where the semicolon denotes the covariant derivative.

Observations have shown that the current universe is very close to a spatially flat geometry \cite{spergel2007three, aghanim2020planck, alam2021completed}. Therefore, in this paper, we shall consider a homogeneous and isotropic Friedman-Lemaitre-Robertson-Walker (FLRW) universe with flat spatial sections, described by the Robertson-Walker metric:
$$ds^{2}=dt^{2}-a(t)^2\delta_{ij}dx^idx^j,$$
where $a(t)$ is the cosmological scale factor, and the Latin indexes run over three-space.

To effectively incorporate NLED into a homogeneous and isotropic geometry framework, employing an averaging technique that satisfies specific criteria \cite{tolman1930temperature}. These criteria include ensuring that the volumetric average of the electromagnetic field remains direction-independent \cite{tolman1930temperature}, that field fluctuations are equally probable in all directions \cite{lemoine1995fluctuations, lemoine1995primordial}, and that there is no net energy flow as observed by comoving observers. Additionally, it is assumed that the electric and magnetic fields, as random fields, possess coherent lengths much shorter than cosmological horizon scales. This ensures that the NLED equations are compatible with the FLRW geometry, facilitating a consistent analysis of cosmological dynamics \cite{de2002nonlinear, novello2004nonlinear, novello2007cosmological, garcia2014no, garcia2014comment}.

Under these conditions, the average energy-momentum tensor adopts the perfect fluid form:
\begin{equation}
    \left\langle T_{\mu\nu}^{EM}\right\rangle=(\rho_{EM}+p_{EM})\frac{u_\mu u_\nu}{c^2}-p_{EM} g_{\mu \nu},
\end{equation}
where the density and pressure of the nonlinear radiation are given by:
\be
    \rho_{EM}=-L+G L_{,G}-4L_{,F} E^{2}, \;\;\;\;\;\;\;\; p_{EM}=L-G L_{,G}-\frac{4}{3}\left( 2B^{2}-E^{2}\right) L_{,F},\nonumber
\ee 
where $E^2$ and $B^2$ are the averaged electric and magnetic fields squared.

In cosmological models coupled with NLED, particularly in the context of the early universe, the analysis often focuses solely on magnetic fields, setting the electric field to zero ($E = 0$). This simplification stems from several compelling reasons. Firstly, the early universe's hot, dense plasma environment favors magnetic fields due to magnetohydrodynamic effects and their energy density scaling with $B^2$ compared to $E^2$ \cite{lemoine1995primordial}. Secondly, limiting the analysis to magnetic fields reduces the complexity of the dynamical equations. Thirdly, NLED can give rise to intriguing magnetic phenomena, such as magnetic monopoles or the self-organization of magnetic structures, making magnetic fields a central focus for understanding these phenomena and their potential cosmological implications. Finally, in many cosmological scenarios, the effects of the electric field are negligible compared to those of magnetic fields, particularly in the early universe. This approach, commonly referred to as studying magnetic universes (MU), enables the investigation of the specific effects of NLED on magnetic phenomena in the early universe while maintaining a manageable analytical framework \cite{de2002nonlinear, novello2004nonlinear, novello2007cosmological, garcia2014no, garcia2014comment}. Thus:
\be
    \rho_B=-L,\;\;\;\;\;\;\;\;p_B=L-\frac{4}{3}FL_{,F},\label{rhoB-pB}
\ee 
and $F=2B^2$ and $G=0$.

In the context of a MU, where NLED is incorporated, the Friedmann equations describe the dynamics of the universe. These equations can be written for the energy density and pressure components, considering the contribution of a magnetic field. The total energy density $\rho_t$ and pressure $p_t$ in such a universe include contributions from both matter and the nonlinear electromagnetic field. 

We are now ready to formulate the dynamical equations for the FLRW model coupled with dark matter and NLED. Even this simplified framework can provide significant physical insights. We shall focus on MU driven by electromagnetic Lagrangian densities that depend only on the invariant $F=2B^2$. In this case, the cosmological equations take the following form:

\bea
    3H^2 &=& \rho_m + \rho_B = \rho_m - L, \nonumber \\
    2\dot{H} &=& -(\rho_m + p_m) - (\rho_B + p_B) = -(\rho_m + p_m) + \frac{4}{3}F L_{,F}, \nonumber \\
    \dot{\rho}_m &=& -3H \rho_m (1 + \omega_m), \quad\quad\quad\quad\quad\quad \dot{F} = -4H F, \label{feqs}
\eea
where $H = \dot{a}/a$ is the Hubble parameter (the overdot denotes a derivative with respect to cosmic time $t$), and $\omega_m$ is the barotropic parameter of the equation of state of the ordinary matter.

The solutions to the last equations in (\ref{feqs}) are given by:

\begin{equation}
    \rho_m = \frac{\rho_{0m}}{a^{3(1 + \omega_m)}}, \quad\quad\quad\quad\quad F = \frac{F_0}{a^4}, \label{solfeqs}
\end{equation}
where $F_0$ and $\rho_{0m}$ are integration constants, and $a = a(t)$ is the scale factor.

\subsection{Cosmological parameters}

This section provides an in-depth introduction of cosmological parameters that play a fundamental role in our analysis \cite{dodelson2020modern}. These parameters allow us to characterize the different critical points that emerge from the analysis using dynamical systems. These critical points represent the initial conditions from which all trajectories of cosmological model evolution originate, the points toward which these trajectories inevitably converge during their evolution, and the transient stages through which the models pass on their evolution. Understanding these parameters provides insights into various features that can be tested against astronomical observations.

\subsubsection{Energy density parameter}

The energy density parameter, denoted as $\Omega$, is a dimensionless quantity that plays an important role in cosmology. It is defined as the ratio of the energy density $\rho_x$ of a particular component $x$ (such as matter, radiation, or dark energy) to the critical density $\rho_{crit}$, which is the density required for the universe to be spatially flat. Mathematically, it is expressed as:
\be
    \Omega_x = \frac{\rho_x}{\rho_{crit}}, \;\;\;\;\; \rho_{crit} = 3H^{2}, \label{OmegaDef}
\ee
where $H$ is the Hubble parameter.

The energy density parameter allows cosmologists to characterize the universe's composition and predict its dynamics. It provides insights into whether the universe is open ($\Omega < 1$), closed ($\Omega > 1$), or flat ($\Omega = 1$).

The energy density parameter is essential for interpreting observational data, evaluating the stability of cosmological models, and understanding the universe's ultimate fate. It helps identify the relative contributions of different components, such as dark matter, dark energy, and ordinary matter, to the overall energy budget of the universe.

\subsubsection{Barotropic parameter}

The barotropic parameter (BP), $\omega$, is a key relationship in cosmology and fluid physics that describes the connection between the pressure $p$ and the energy density $\rho$ of a perfect fluid: 
\be
    \omega = \frac{p}{\rho}. \label{ParBar}
\ee
where $p$ is the pressure of the fluid, and $\rho$ is its energy density. This parameter is important because it determines how the fluid affects the universe's expansion. Depending on the value of $\omega$, the dynamic behavior of the fluid can vary significantly. 

When considering a cosmological model with multiple material components, such as matter, radiation, and nonlinear electromagnetic fields, the effective barotropic parameter $\omega_{eff}$ provides a useful way to describe the overall relationship between the total pressure and the total energy density of the universe. The effective barotropic parameter is defined as:
\be
    \omega_{eff} = \frac{p_{t}}{\rho_{t}}, \label{ParBar_eff}
\ee
$p_{t}$ and $\rho_{t}$ are the total pressure and energy density, respectively, summed over all components. 

The effective barotropic parameter $\omega_{eff}$ can also be expressed in terms of the Hubble parameter $H$ as follows: 
\be
\omega_{eff} = -1 - \frac{2\dot{H}}{3H^2}. \label{weffH}
\ee

This expression provides a direct relationship between the effective barotropic parameter and the Hubble parameter, offering insights into the overall dynamics of the universe.

\subsubsection{Deceleration parameter}

The deceleration parameter, denoted by $q$, is a dimensionless quantity that describes the rate of change of the universe's expansion rate. It indicates whether the expansion of the universe is accelerating or decelerating. The deceleration parameter is defined as:
\be
    q = -\frac{\ddot{a} a}{\dot{a}^2}, \label{q}
\ee
where $a$ is the scale factor and $\dot{a}$ and $\ddot{a}$ are the first and the second time derivative of $a$.

The deceleration parameter can be expressed in terms of the Hubble parameter $H$ and its time derivative $\dot{H}$ as follows: 
\be
    q = -1 - \frac{\dot{H}}{H^2}. \label{qH}
\ee

The deceleration parameter is essential for understanding the universe's expansion history and future evolution.

\subsubsection{Squared sound speed}

The SSS, denoted as $c_s^2$, is a fundamental parameter in cosmology and fluid dynamics that describes the propagation speed of pressure (acoustic) waves through a given medium. It is defined as the ratio of the change in pressure $p$ to the change in energy density $\rho$ and is mathematically expressed as \cite{peebles2003cosmological, ellis2007causality}:
\be
    c_s^2 = \frac{dp}{d\rho}. \label{cs2}
\ee

The SSS is important for several reasons. The value of $c_s^2$ helps determine the stability of a cosmological model or a fluid. For a stable system, $c_s^2$ must be non-negative ($c_s^2 \geq 0$) \cite{yang2012dynamics}. A negative SSS indicates instability, leading to exponential growth of perturbations.

Even if $c_s^2$ is a positive quantity, a causality issue may arise whenever the squared sound speed exceeds the local speed of light \footnote{For a critical review on this issue, see \cite{ellis2007causality}}. Indeed, it is commonly assumed that $c_s^2 \leq 1$, while the complementary bound $c_s^2 > 1$ is employed as a criterion for rejecting theories \cite{hawking1973large, wald2010general}.
    
In the following section, we will analyze the stability and causality conditions of several NLED Lagrangians widely studied in the literature.


\section{Stability and causality analysis of several NLED Lagragians}

For a Lagrangian that only depends on the invariant $F$ and thus corresponds only to magnetic fields where $E = 0$, as is the case in this work, the adiabatic speed of sound squared (SSS) for scalar perturbations can be expressed as:
\be 
    c_s^2=\frac{dp_B/dF}{d\rho_B/dF}=\frac{1}{3}+\frac{4}{3}\frac{F L_{,FF}}{L_{,F}}, \label{speed-sound}
\ee 
where $p_B$ and $\rho_B$ are given by Eqs. (\ref{rhoB-pB}). Hence, the above-discussed bounds on the SSS, $0\leq c^2_s\leq 1$, translate into the following bounds on the NLED Lagrangian and its $F$-derivatives:
\be 
    -\frac{1}{4}\leq\frac{FL_{,FF}}{L_{,F}}\leq\frac{1}{2}.\label{c2s-bounds}
\ee

Following the causality principle, the group velocity of excitations over the background should be less than the speed of light, ensuring no tachyons are in the theory spectrum. The unitarity principle guarantees the absence of ghosts. Both principles lead to the inequalities \cite{shabad2011effective}:
\be
    L_F \leq 0, \;\;\;\; L_{FF} \geq 0, \;\;\;\; L_F+2FL_{FF} \leq 0,  \label{CUC}
\ee 
for a Lagrangian that only depends on the invariant $F$.

As we can see, the stability and causality of the material content of the model corresponding to NLED radiation depend strongly on its Lagrangian dependence on the electromagnetic invariant $F$.

\subsection{Power-Law NLED Lagrangian}\label{sec-pow-law}

Magnetic universes have been extensively explored within the framework of NLED theories, often characterized by simple Lagrangian densities. One of the simplest Lagrangians is given by:
\be
    L=-\frac{1}{4}F+\alpha F^2,\label{LF2}
\ee
where the nonlinear term $\propto F^2$ \cite{de2002nonlinear, camara2004nonsingular, camara2007nonlinearity, campanelli2008inflation, vollick2008homogeneous, de2009classification, garcia2014no} has been suggested to potentially induce a cosmic bounce, thus avoiding the initial singularity known as the big bang \cite{de2002nonlinear}. On the other hand, Lagrangians featuring inverse powers of the electromagnetic field $F$ are intriguing due to the potential significance of nonlinear electromagnetic effects in both the early and late stages of cosmic evolution. Models employing Lagrangian densities like \cite{novello2004nonlinear, novello2007cosmological, garcia2014no}:
\be
    L=-\frac{1}{4}F-\frac{\gamma}{F},\label{LF-1}
\ee
have been proposed to elucidate the late-time accelerated expansion of MU \cite{novello2004nonlinear}. Additionally, combinations of positive and negative powers of $F$ have been investigated \cite{novello2007cosmological, maity2011correspondence, garcia2014no}, as exemplified by:
\be
    L=-\frac{1}{4}F-\frac{\gamma}{F}+\alpha F^2.\label{LF2F-1}
\ee

This composite model captures key cosmic evolutionary stages and remarkably avoids the cosmological Big Bang singularity. Specifically, the quadratic term $\propto F^2$ dominates during early epochs, facilitating a nonsingular bounce, while the Maxwell term $\propto -F$ takes precedence in the radiation era. At late times, the term $\propto F^{-1}$ governs, driving cosmic acceleration \cite{novello2007cosmological}.

In \cite{garcia2014no}, it was demonstrated that numerous cosmological models rooted in NLED Lagrangians (\ref{LF2})-(\ref{LF2F-1}) are susceptible to curvature singularities of sudden and/or big rip varieties, or exhibit pronounced instability against minor perturbations of the cosmological background—often attributed to the negative sign of SSS. Moreover, concerns regarding causality may emerge due to the potential superluminal propagation of background perturbations.

The subsequent extension of the Lagrangian density involves incorporating a power law dependence on the invariant $F$, expressed as:
\be
    L=-\gamma F^\alpha.\label{FFm1agamma}
\ee

In this scenario, upon evaluating the SSS (Eq. \ref{speed-sound}), we find $c_s^2=-1+\frac{4}{3} \alpha$. It becomes evident that the Lagrangian density of NLED remains causal and stable within the range of $\alpha$ values spanning $3/4 \leq \alpha \leq 3/2$.

Now, let us examine the Lagrangian density of NLED expressed as:
\be
    L=-\frac{1}{4}F-\gamma F^\alpha. \label{FFm1a}
\ee

This formulation characterizes a universe featuring linear Maxwell radiation alongside a nonlinear power law term. A prior investigation \cite{garcia2014no} explored specific cases of $\alpha$ ($\alpha=-1$ and $\alpha=2$), revealing that these models fail to meet the necessary constraints for SSS. In \cite{montiel2014parameter} examined a variant incorporating cold dark matter ($\rho_m$ with $p_m=0$). Their analysis, anchored in astronomical observations and with $\gamma < 0$, demonstrated that the Lagrangian (\ref{FFm1a}) accurately replicates dark energy phenomena for $\alpha = -1/4$ and $\alpha=-1/8$. More recently, \cite{joseph2021cosmology} explored a cosmological framework involving a variable Newton's constant, $G(t)$, alongside a NLED of the form $L=-\frac{F^\alpha}{4}$. They established its stability within $5/2 \geq \alpha \geq 7/4$.

The expression for the SSS, given by Eq. (\ref{speed-sound}), corresponding to the Lagrangian (\ref{FFm1a}) in terms of the scale factor $a$ and the parameter $\alpha$, is as follows:
\be
    c_s^2=\frac{1}{3}+\frac{16 \alpha (\alpha-1) \gamma F^{\alpha}}{12 \alpha \gamma F^{\alpha}+3F}=\frac{1}{3}+\frac{16 \alpha (\alpha-1) \gamma a^{4-4 \alpha}}{12 \alpha \gamma a^{4-4 \alpha}+3}. \label{SSSPL}
\ee
where we have used the solution to the equation for the conservation of the electromagnetic field (\ref{solfeqs}).

\begin{figure}[H]
    \centering
    \includegraphics[width=9cm]{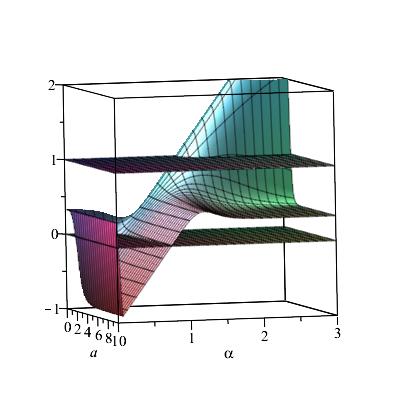}
    \includegraphics[width=7cm]{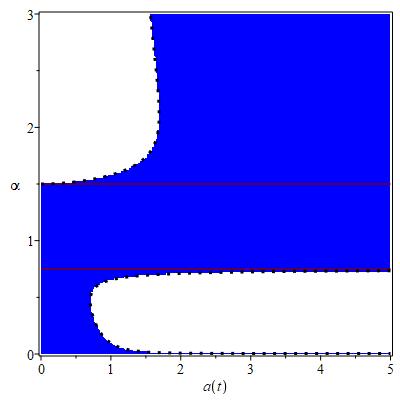}
    \caption{The graph shows the behavior of the speed of sound squared for the Lagrangian (\ref{FFm1a}). In the left panel, we show the surface of $c_s^2$ with respect to the scale factor $a$ and the parameter $\alpha$. The lines $\alpha=3/4$ and $\alpha =3 /2$ are in the right panel. Here we are considering $F_0=1$.} \label{Palpha}
\end{figure}

The plot on the right panel of Fig. \ref{Palpha} shows that the SSS remains stable and causal within a narrow range of $\alpha$ values, specifically $3/4 \leq \alpha \leq 3/2$ and $\alpha=0$ for the Eq. (\ref{SSSPL}). In \cite{montiel2014parameter}, the NLED Lagrangian (\ref{FFm1a}) was considered, and for the obtained values of $\alpha$ (namely, $\alpha=-1/4$ and $\alpha = -1/8$), the resulting SSS were $c_s^2=-4/3$ and $c_s^2=-7/6$, respectively, indicating instability in their model. Similarly, the cases with $\alpha=-1$ and $\alpha=2$, as studied in \cite{garcia2014no}, yielded $c_s^2=-7/3$ and $c_s^2=5/3$, respectively, reaffirming the instability and non-causality of these models.

Based on the findings above, we can deduce that the NLED Lagrangian density represented by (\ref{FFm1a}) demonstrates causality and stability within the range $3/4 \leq \alpha \leq 3/2$, concerning minor perturbations of the background. Our subsequent endeavor involves a dynamical analysis to unveil the evolution of the cosmological model integrating ordinary matter $\rho_m$ and nonlinear radiation $\rho_B$. This examination will be detailed in the subsequent section.

\subsection{Generalized rational nonlinear electrodynamics}

In this section, we explore a generalized model of NLED described by the following Lagrangian density:

\be 
    L = -\frac{F}{4} - \frac{b F}{1 + \epsilon (2 \epsilon \beta F)^\alpha}, \label{GRNLED}
\ee
where $b$, $\epsilon = \pm 1$, and $\alpha$ are a dimensionless parameter, and $\beta$ is a parameter with dimensions $[L]^4$. To ensure that the Lagrangian remains real, in \cite{kruglov2023nonlinear} states that $\epsilon = 1$ for $B > E$ and $\epsilon = -1$ for $B < E$. However, in our case, since $E = 0$, we will only consider the scenario where $\epsilon = 1$.  We recover Maxwell's electrodynamics when $b = 0$. 

Furthermore, when we apply the weak field approximation ($\beta F \ll 1$), the power series expansion of $L$ (Eq. (\ref{GRNLED})) reveals that nonlinearity persists. This implies the existence of nonlinearity even at late times, which is important for exploring its implications on the model's behavior during those periods, providing this Lagrangian density proves to be stable and causal.

Some particular cases of the Lagrangian density (\ref{GRNLED}) have been previously analyzed in various studies \cite{kruglov2015universe, kruglov2015model, kruglov2017nonlinear, mazharimousavi2019electric, mazharimousavi2019note, kruglov2022nonlinearly}, primarily investigating black hole solutions with different values of $\alpha$, particularly when $\alpha = 1/2$, and for both values of $\epsilon$. In \cite{kruglov2023nonlinear}, it was demonstrated that singularities of point electric charges are absent, and the electromagnetic energy is finite.

The expression for the SSS, given by Eq. (\ref{speed-sound}), corresponding to the Lagrangian (\ref{GRNLED}) in terms of the scale factor $a$ and the parameters $\alpha$, $\beta$, and $b$, is as follows:
\be
    c_s^2=\frac{1}{3}-\frac{16}{3} \frac{b \alpha \left[(\alpha-1) (2 \beta F)^{2\alpha}-(\alpha+1)(2 \beta F)^\alpha \right]}{(1+ (2 \beta F)^\alpha) \left[ (2-4b(\alpha-1)) (2 \beta F)^\alpha+(2\beta F)^{2 \alpha}+4 b+1 \right]}. \label{SSSGRN}
\ee

The analysis to determine the parameter values for which this Lagrangian density is stable and causal is quite complex. Therefore, we will start by analyzing some specific cases. Initially, we will consider the model without the linear Maxwell term and subsequently include it in our analysis.

\subsubsection{Generalized Rational Nonlinear Electrodynamics Without Maxwell Term}

We will examine in more detail a generalized model of rational nonlinear electrodynamics \cite{kruglov2024nonlinear}, excluding the linear Maxwell term and focusing on magnetic universes, described by the following Lagrangian density ($\epsilon=1$):

\be 
    L = -\frac{bF}{1 + (2 \beta F)^\alpha}, \label{NLEDG}
\ee 

In the limit as $\beta \to 0$, the Lagrangian (\ref{NLEDG}) reduces to Maxwell's theory. Specific cases of this Lagrangian density have been studied in the cosmological context \cite{kruglov2015model, kruglov2024nonlinear}, detailed in \cite{kruglov2015universe, otalora2018inflation, singh2018accelerating, kruglov2017inflation, kruglov2024universe}.

The squared sound speed for this Lagrangian density is given by:

\be 
    c_s^2 = \frac{1}{3} - \frac{4}{3} \frac{[(\alpha-1)(2 \beta F)^{2\alpha} - (2 \beta F)^\alpha(1+\alpha)] \alpha}{[1 + (2 \beta F)^\alpha][(\alpha-1)(2 \beta F)^\alpha-1]}.
    \label{cs2NLEDG}
\ee 

As we can see, the expression for the squared sound speed does not depend on the parameter $b$, so it will not influence stability or causality. However, it may affect the dynamics of the model. We will plot the SSS expressed in Eq. (\ref{cs2NLEDG}) for some $\beta$ parameter values to determine whether this SSS remains stable and causal.

\begin{figure}[H]
    \centering
    \includegraphics[width=9cm]{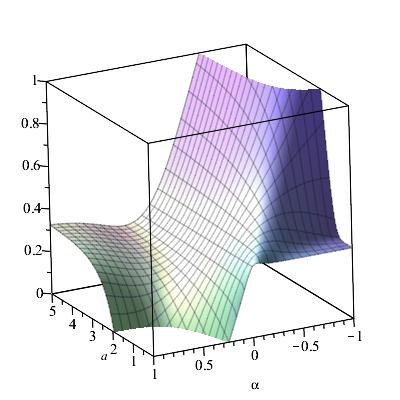}
    \includegraphics[width=7cm]{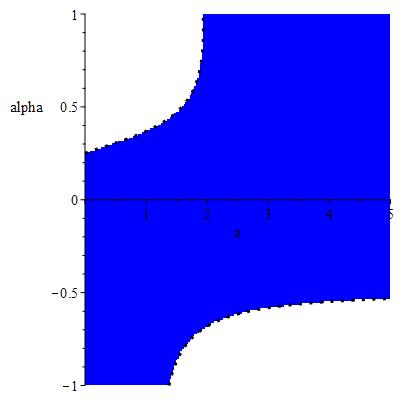}
    \caption{The left panel depicts the behavior of SSS for specific values of $\alpha$ (Eq. (\ref{cs2NLEDG})) where it exhibits causality and stability.  In the right panel, the blue region indicates the causal and stable SSS conditions, corresponding to the region between the lines $\alpha=-1/2$ and $\alpha =1/4$. We are considering $F \propto a^{-4}$ and $\beta = 1$.}
    \label{cs2e1}
\end{figure}

As illustrated in the right panel of Figure \ref{cs2e1}, the blue region represents where the SSS is both causal and stable. The right panel also depicts the behavior of SSS, revealing its consistent tendency towards linear Maxwell radiation in the late-time regime. This model warrants further investigation through a dynamical systems analysis, particularly within the parameter range $-\frac{1}{2} \leq \alpha \leq \frac{1}{4}$, where the SSS stability and causality.

In the figure, we have used $\beta = 1$. However, changing this positive value shifts the regions of instability and acausality closer to or further from the origin where $a = 0$. This adjustment does not affect the interval of the $\alpha$ parameter where $0 \leq c_s^2 \leq 1$.

Recently, in \cite{kruglov2024universe}, the dynamics of this model have been analyzed. However, that work's stability and causality analysis were verified only with values outside the range we found here. Therefore, in our work, we can perform a more comprehensive dynamical analysis within the parameter space we have identified as stable and causal.

From the above, we can infer a specific region in the parameter space for the Lagrangian (\ref{NLEDG}), where a dynamical analysis can be conducted in a homogeneous and isotropic framework to observe its effects during early and late times.

Another specific case of the Lagrangian density (\ref{NLEDG}) is given by:
\be 
    L = -\frac{bF}{1 + 2\beta F}, \label{NLED10}
\ee 
where $\alpha = 1$, which has been studied in \cite{kruglov2015model, kruglov2015universe, otalora2018inflation, singh2018accelerating, kruglov2020rational}.

The SSS (Eq. \ref{speed-sound}) for this Lagrangian density (\ref{NLED10}) is given by:
\be 
    c_s^2 = \frac{1 - 14\beta F}{3 + 6\beta F} = \frac{a^4 - 14\beta}{3a^4 + 6\beta},
    \label{cs2betaFb}
\ee
where we have used the solution for the conservation equation of the magnetic field, $F \propto a^{-4}$.

As we can see, the SSS does not depend on the parameter $b$, similar to the previous case. If we take the limit as $a \to 0$, then $c_s^2 \to -\frac{7}{3}$, while if $a \to \infty$, $c_s^2 \to \frac{1}{3}$, for all values of $\beta > 0$. This implies that the model begins unstable, becomes stable at some point ($0 \leq a \leq 14^{1/4} \beta^{1/4}$ it is unstable), and eventually ceases to be causal \cite{kruglov2015nonlinear, singh2018accelerating}. The value of the scale factor $a$ at which the model becomes stable depends on $\beta$, allowing us to adjust this point closer to or farther from $a = 0$. In any other case, there will always be an initial period of instability \cite{kruglov2015universe}. Consequently, this NLED model can be ruled out as a candidate for describing the current universe throughout its evolutionary range.

\subsubsection{Generalized Rational Nonlinear Electrodynamics With Maxwell Term}

In this subsection, we investigate the general scenario by adding Maxwell's linear electrodynamics term (Eq. \ref{GRNLED}) to observe the behavior of the SSS. However, due to the complexity of performing this analysis generally, we focus on some specific cases.

The expression for the SSS, obtained by substituting the Lagrangian (Eq. (\ref{GRNLED})) into Eq. (\ref{speed-sound}), is given by Eq. (\ref{SSSGRN}). This SSS depends on the parameter $b$, which influences the evolution of a cosmological model with this material component during both early and late times. 

\begin{figure}[H]
    \centering
    \includegraphics[width=9cm]{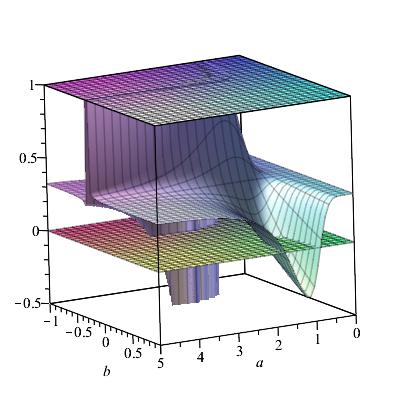}
    \includegraphics[width=7cm]{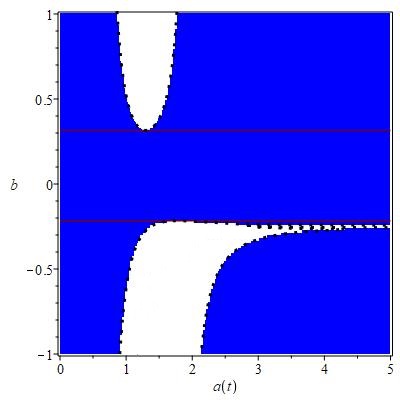}
    \caption{The left panel shows $c_s^2$ in terms of the scale factor $a$ and the parameter $b$. In the right panel, we show the region in blue where $0 \leq c_s^2 \leq 1$ for the Lagrangian (\ref{RNLED1}). As shown, a small range of $b$-values for which the $c_s^2$ remains within the range $0\leq c^2_s\leq 1$, $-\frac{27}{125} \leq b \leq \frac{108}{343}$.}
    \label{F3}
\end{figure}

The first particular case of this Lagrangian density (\ref{GRNLED}) when $\alpha=1$ is given by \cite{kruglov2015model}:
\be
    L=-\frac{F}{4}-\frac{bF}{1+2\beta F}. \label{RNLED1} 
\ee 
Here, the SSS for this Lagrangian density is given by:
\be
    c_s^2 = \frac{1}{12} \frac{8 + (4b + 1)a^{12} - 56ba^8 + 6a^8 + 12a^4}{(1 + (b + 1/4)a^8 + a^4)(a^4 + 2)}, \label{cs2z}
\ee
where we are considering $F \propto a^{-4}$. We can find an interval of values of parameter $b$ in which the model is stable and causal ($0\leq c^2_s\leq 1$) given by $-\frac{27}{125}=-0.216 \leq b \leq 0.314= \frac{108}{343}$ as we can see in the left panel of Figure \ref{F3}.

Thus, we can perform the dynamical analysis of this model for values of b within the range we have obtained, where the square sound speed tells us that the nonlinear radiation is causal and stable. In the following section, we will perform this analysis using the dynamical systems tools.

To advance the stability and causality analysis of the SSS given by (\ref{SSSGRN}), we should consider specific values for the parameter $\alpha$ or the parameter $b$, as a general analysis is very complicated. Therefore, we will defer the review of stability and causality to the specific values determined by the dynamic analysis for the Lagrangian density (\ref{GRNLED}) that we will perform in the next section.

\section{Dynamical Analysis of Models with Stable and Causal Lagrangian Density}

In this section, we perform the dynamical analysis of models where nonlinear radiation is stable and causal, as identified in the previous section. We utilize the tools provided by dynamical systems theory for this.

\subsection{Power Law Lagrangian}

In the previous section, we found that a power law model of nonlinear electrodynamics (\ref{FFm1a}) is stable and causal as long as $3/4 \leq \alpha \leq 3/2$. We will now analyze the evolution of a homogeneous and isotropic model whose matter content is given by this NLED and ordinary matter (cold dark matter).

Thus, if we take the density for ordinary matter as $\rho_m$ (considering non-relativistic dust ($\omega_m=0$) and the non-linear electromagnetic field given by the Lagrangian (\ref{FFm1a}), the dynamical equations to study are the following (\ref{feqs}):

\bea
    3H^2 &=& \rho_m + \frac{F}{4} + \gamma F^\alpha, \quad \quad \quad  -2\dot{H} = \rho_m + \frac{1}{3}F + \frac{4}{3}\gamma \alpha F^\alpha, \nonumber \\
    &&\dot{\rho}_m = -3H\rho_m(1+\omega_m), \quad \quad \quad  \dot{F} = -4HF. \label{feq-c1}
\eea 

To use dynamical systems theory, we must transform this second-order system into a first-order system by changing variables:

\be 
    x = \frac{F}{12H^2} = \Omega_r, \quad \quad \quad \quad y = \gamma \frac{F^\alpha}{3H^2} = \Omega_{NL}, \label{var-c1}
\ee 

Thus, the first of the equations (\ref{feq-c1}) can be written as:
\be
    \Omega_m = 1 - x - y, \label{Omxy1}
\ee 
where we see that the variables are bounded within the phase space:
\be 
    \Psi = \{(x, y) \mid 0 \leq x + y \leq 1, 0 \leq x \leq 1, 0 \leq y \leq 1\}. \label{phsp-c1}
\ee 

The second equation (\ref{feq-c1}) takes the form:
\be 
    -2\frac{\dot{H}}{H^2} = 3\Omega_m + 4x + 4\alpha y = 3 + x + (4\alpha - 3)y. \label{raycha-c1}
\ee

Thus, the dynamical system takes the form:
\bea 
    x' &=& -2x(2+\frac{\dot{H}}{H^2})=x[x - 1 + (4\alpha - 3)y], \nonumber \\
    y' &=& -2y(2\alpha+\frac{\dot{H}}{H^2})=y[(4\alpha - 3)y + x + 3 - 4\alpha], \label{dyn-sys-c1}
\eea
where the prime denotes the derivative with respect to $\tau$, which is the conformal time, $\tau = \int \frac{dt}{a(t)}$.
In terms of the variables (\ref{var-c1}) of the phase space, the squared sound speed (\ref{SSSPL}) can be written as:
\be
    c^2_s = \frac{x + \alpha(4\alpha - 3)y}{3(x + \alpha y)}. \label{cs2xy}
\ee

Finally, both the effective barotropic parameter, $\omega_{eff}$, in Eq. (\ref{weffH}) and the deceleration parameter $q$ in Eq. (\ref{qH}) in the new variables are given by:
\be
    \omega_{eff}=\frac{4}{3} \alpha y +\frac{1}{3}x-y,  \quad \quad \quad \quad q=\frac{1}{2}+2 \alpha y+\frac{1}{2}x-\frac{3}{2}y.
\ee

The solution of this dynamical system (\ref{dyn-sys-c1}) has only three critical points within the phase space:

\begin{itemize}
    \item At the first critical point, $P_m(0,0)$, ordinary matter dominates ($\Omega_m = 1$). The behavior of this critical point, as determined by the eigenvalues $\lambda_1 = -1$ and $\lambda_2 = 3 - 4\alpha$, indicates that it is a future attractor when $\alpha > 3/4$ ($c_s^2$-stable) and a saddle point when $\alpha < 3/4$ ($c_s^2$-unstable). The deceleration parameter at this point is $q = 1/2$, corresponding to a decelerated point with an effective barotropic parameter of $\omega_{eff} = 0$.
    
    \item The second point, where Maxwell radiation dominates ($P_M(1,0)$) with $c_s^2 = \frac{1}{3}$, has eigenvalues $\lambda_1 = 1$ and $\lambda_2 = 4 - 4\alpha$. This indicates that it is a source (past attractor) when $\alpha < 1$ and a saddle point when $\alpha > 1$. The deceleration parameter at this point is $q = 1$, and its effective barotropic parameter is $\omega_{eff} = 1/3$.

    \item Finally, at the point where nonlinear radiation dominates ($P_{NL}(0,1)$), the squared sound speed is given by $c_s^2 = \frac{4}{3}\alpha - 1$. The eigenvalues $\lambda_1 = 4\alpha - 3$ and $\lambda_2 = 4\alpha - 4$ indicate that this point is a future attractor if $\alpha < 3/4$ ($c_s^2$-unstable), a saddle point if $3/4 < \alpha < 1$ ($c_s^2$-stable), and a source (past attractor) if $c_s^2$-stable. The deceleration parameter at this point is $q = -1 + 2\alpha$, and its effective barotropic parameter is $\omega_{eff} = -1 + \frac{4}{3}\alpha$.
\end{itemize}

\begin{figure*}[t!]
    \includegraphics[width=8cm]{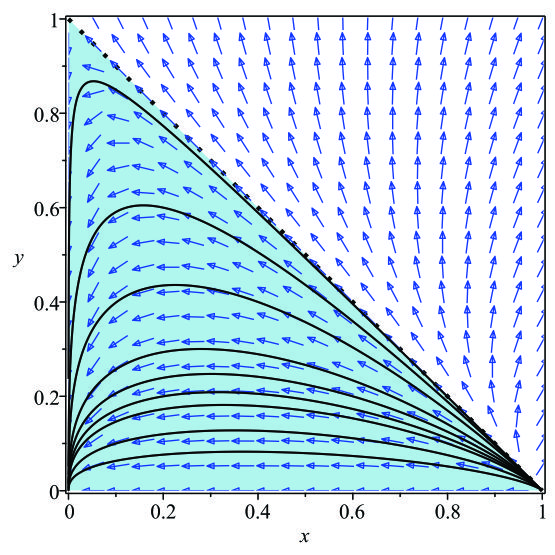}
    \includegraphics[width=8cm]{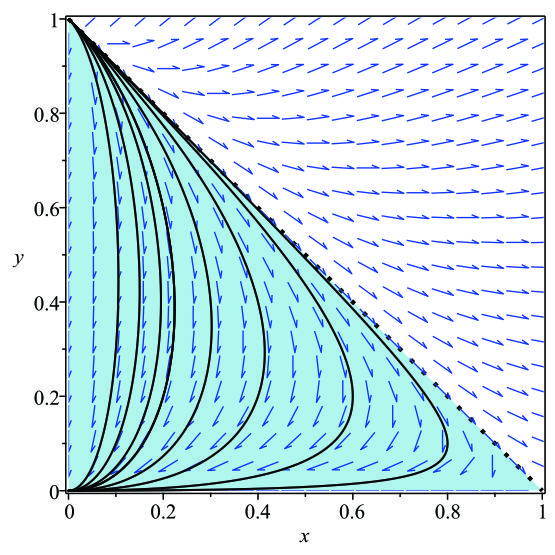}
    \caption{Phase portrait of the dynamical system (\ref{dyn-sys-c1}). In the left panel, $\alpha = 0.85$ (within the interval $3/4 \leq \alpha < 1$), while in the right panel, $\alpha = 1.25$ (within the interval $1 < \alpha < 3/2$). In both cases, the stability and causality conditions ($0 \leq c^2_s \leq 1$) are satisfied, making the models feasible with these values of $\alpha$.} 
    \label{F2}
\end{figure*}

The evolution of the model can be divided into three scenarios depending on the value of the $\alpha$ parameter:

\begin{itemize}
    \item For $\alpha < 3/4$ or $\alpha > 3/2$, the model starts from a Maxwell radiation domain ($P_M$), transiently passes through a matter-dominated phase ($P_m$), and finally reaches a nonlinear radiation domain ($P_{NL}$). This scenario is not physically acceptable because the radiation is unstable and acausal.

    \item For $3/4 \leq \alpha < 1$, all trajectories start from the Maxwell radiation-dominated point ($P_M$). If $\alpha$ is close to $3/4$, trajectories transiently approach the nonlinear radiation-dominated point ($P_{NL}$) but eventually reach the matter-dominated point ($P_m$). If $\alpha$ is closer to $1$, the trajectories decay almost directly to the matter domain point ($P_m$), as shown in the left panel of Fig. \ref{F2}.

    \item For $1 < \alpha < 3/2$, the behavior of the phase space is slightly different. The trajectories begin at the nonlinear radiation domain point ($P_{NL}$), evolve transiently near the Maxwell radiation-dominated point ($P_M$), and finally reach the matter-dominated point ($P_m$), as shown in the right panel of Fig. \ref{F2}.
\end{itemize}  

In this section, we have analyzed the dynamical evolution of a power law model of nonlinear electrodynamics, as given by the Lagrangian density (\ref{FFm1a}). This model was previously shown to be stable and causal within $3/4 \leq \alpha \leq 3/2$. We identified critical points and evaluated their stability by transforming the second-order equations into a first-order autonomous system. Our findings reveal three distinct scenarios depending on the parameter $\alpha$ value. The phase space exhibits two stable and causal behaviors depending on the value of $\alpha$. In one scenario, the model transitions from a Maxwell radiation-dominated state to a nonlinear radiation-dominated state and finally to an ordinary matter-dominated state. In the other scenario, the model transitions from a nonlinear radiation-dominated state to a Maxwell radiation-dominated state and ultimately to an ordinary matter-dominated state. These results provide a comprehensive understanding of the dynamical evolution and stability of the power law nonlinear electrodynamics model.

\subsection{Generalized rational nonlinear electrodynamics dynamical evolution without Maxwell term} \label{ssgrnled}

The next model we will analyze using dynamical systems corresponds to the Lagrangian density given by (\ref{NLEDG}), which we have demonstrated to be causal and stable for the range $-1/2 \leq \alpha \leq 1/4$. The dynamical equations for this model are given by:
\be
    3H^2=\rho_m+\frac{bF}{1+(2\beta F)^\alpha}, \quad -2\dot H=\rho_m-\frac{4bF}{3}\Big[\frac{(\alpha-1)(2\beta F)^\alpha-1}{(1+ (2\beta F)^\alpha)^2}\Big], \label{feq-c2alpha}
\ee and $\dot{F} = -4H F$. Here, we consider ordinary matter $\rho_m$ (non-relativistic dust $\omega_m=0$).

We consider the following new variables to obtain an autonomous first-order dynamical system:
    
\be 
    x=\frac{F}{3H^2},\;\;\;\;\;\;\;\; 
    y=\frac{(2\beta F)^\alpha}{1+(2\beta F)^\alpha}. \label{newvarsG}
\ee 

The phase space is defined by $\Psi = \{(x, y) \mid 0 \leq x \leq 1, \; 0 \leq y \leq 1\}$. Thus, the dynamical equations (\ref{feq-c2G}) in these new variables can be expressed as:

\be
    \Omega_m=1-bx(1-y), \;\;\;\;\;\;\;\; -2\frac{\dot H}{H^2}=3\Omega_m-4bx(\alpha y-1)(1-y),\label{feq-c2G}
\ee
where $\Omega_m=\frac{\rho_m}{3H^2}$. 

The deceleration parameter, effective barotropic parameter, and SSS (\ref{cs2NLEDG}), in terms of the new variables, are given by:
\bea
    q&=&\frac{1-bx(1-y)(4\alpha y-1)}{2}, \nonumber \\
    \omega_{eff}&=&-\frac{bx(4\alpha y-1)(1-y)}{3}, \nonumber \\
    c_s^2&=&\frac{(-8y^2+4y) \alpha^2+5y\alpha-1}{3\alpha y-3}.
\eea

The 2D dynamical system for this model, described by the equations (\ref{feq-c2GN}), in terms of the new bounded variables, takes the following form:
\begin{equation}
    x' = -2x(2+\frac{\dot{H}}{H^2})=4b\alpha x^2 y^2-4\alpha b x^2 y-b x^2 y+b x^2-x, \;\;\;\;\;
    y' = -4 \alpha y (1-y), \label{DynSysGRb}
\end{equation}
where the prime denotes differentiation with respect to $\tau = \ln{a}$.

Below, we show the critical points of the dynamical system (\ref{DynSysGRb}). We include the values of several cosmological parameters of observational significance evaluated at the critical points:

\begin{enumerate}
    \item The first critical point, $P_1(0,0)$, dominated by ordinary matter ($\Omega_m = 1$), is decelerated ($q = \frac{1}{2}$), and its effective barotropic parameter simulates dust ($\omega_{eff} = 0$). At this point, the SSS corresponds to Maxwell radiation, meaning $c_s^2 = \frac{1}{3}$. The Jacobian matrix's eigenvalues are $\lambda_1 = -1$ and $\lambda_2 = -4\alpha$. This indicates that $P_1$ is a future attractor if $\alpha > 0$ and a saddle point if $\alpha < 0$.
    
    \item The point $P_2(0,1)$ is also dominated by ordinary matter, is decelerated ($q = \frac{1}{2}$), and has an effective barotropic parameter that simulates dust ($\omega_{eff} = 0$). The SSS at this point is given by $c_s^2 = \frac{1}{3}(1-4\alpha)$. Therefore, the stability and causality depend on the value of $\alpha$, which must satisfy the condition found in the previous section: $-\frac{1}{2} \leq \alpha \leq \frac{1}{4}$. The eigenvalues of the Jacobian matrix are $\lambda_1 = -1$ and $\lambda_2 = 4\alpha$, indicating that $P_2$ is a future attractor if $\alpha < 0$ and a saddle point otherwise.

    \item Finally, the point $P_3\left(\frac{1}{b}, 1\right)$ is dominated by radiation ($\omega_m = 0$) and is decelerated ($q = 1$). Its effective barotropic parameter simulates Maxwell radiation ($\omega_{eff} = \frac{1}{3}$) and the SSS is $c_s^2 = \frac{1}{3}$. The eigenvalues of the Jacobian matrix at this point are $\lambda_1 = 1$ and $\lambda_2 = -4\alpha$, indicating that this point acts as a past attractor if $\alpha < 0$ and as a saddle point for other values.

\end{enumerate}

\begin{figure*}[h]
    \centering
    \includegraphics[width=8cm]{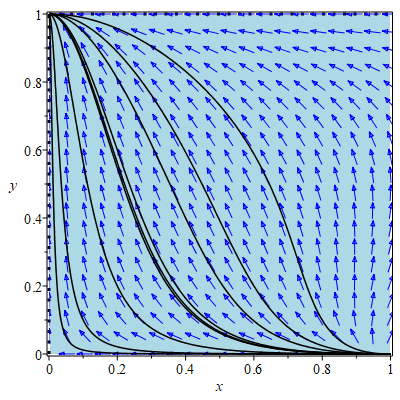}
    \includegraphics[width=8cm]{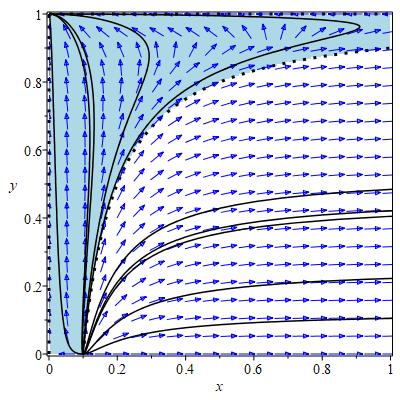}
    \caption{Phase portrait of the dynamical system (\ref{DynSysGRb}) for different parameter values of $\alpha$. In the left panel, it shows the trajectories of the model given by the equations (\ref{DynSysGRb}) with $\alpha = -\frac{1}{2}$ and $b=1$. In the right panel, it is shown for $\alpha = -\frac{1}{2}$ but in this case $b = 10$. The blue region corresponds to the constraint $0 \leq \Omega_m \leq 1$.}
    \label{DynSysGRbdi}
\end{figure*}

For the case when $-1/2 \leq \alpha \leq 0$, the phase space of the cosmological model features a past attractor, a saddle point, and a future attractor. This configuration implies a well-defined evolutionary pathway for the universe, starting from a common initial state represented by the past attractor. The trajectories then pass through a transitional phase governed by the saddle point before converging towards the future attractor. These critical points indicate a structured progression through different dynamical regimes, providing insights into the possible phases of cosmic evolution. The past attractor suggests a dominant initial condition. In contrast, the future attractor represents the universe's ultimate fate under the given model, with the saddle point highlighting intermediate transitional behaviors. Figure (\ref{DynSysGRbdi}) shows the behavior within this interval for two values of the parameter $b$ when $\alpha = -1/2$.

For the case when $0 < \alpha < 1/4$, the phase space of the cosmological model lacks a past attractor, indicating that there is no common initial state from which all trajectories originate. Instead, the system's evolution is characterized by trajectories that do not converge to a single source in the past. This absence of a past attractor implies a less structured and potentially more chaotic initial phase of the universe's evolution. The lack of a unified starting point for the phase trajectories can make predicting the initial conditions and their subsequent evolution challenging. Despite this, the presence of a future attractor still suggests a common endpoint for the universe's evolution. However, the model may be considered less robust or less physically plausible due to the absence of a well-defined initial state. Whether this model is entirely dismissible depends on further analysis of its physical implications and compatibility with observational data.

In this section, we have analyzed the dynamical evolution of the generalized rational nonlinear electrodynamics model given by the Lagrangian density (\ref{NLEDG}). We focused on the parameter range $-1/2 \leq \alpha \leq 1/4$, where the model was previously demonstrated to be causal and stable. We have identified critical points and evaluated their stability by transforming the second-order equations into a first-order autonomous system. Our findings indicate distinct evolutionary paths for different values of $\alpha$. For $\alpha < 0$, the model features a well-defined evolutionary pathway with a past attractor, a saddle point, and a future attractor, suggesting a structured cosmic evolution. Conversely, for $0 < \alpha < 1/4$, the absence of a past attractor implies a more chaotic initial phase, yet the presence of a future attractor provides a common endpoint for the universe's evolution. These results highlight the importance of parameter selection in determining the physical viability and predictive power of cosmological models based on nonlinear electrodynamics.

\subsection{Dynamical Evolution of Rational Nonlinear Electrodynamics with Maxwell term}

In this section, we analyze a stable and causal NLED model for a limited range of values of its parameter $b$ for the Lagrangian density given by Eq. (\ref{RNLED1}), which is a particular case of the Lagrangian density (\ref{GRNLED}) with $\alpha = 1$ \cite{kruglov2015model}. To determine the dynamics of this model within the stable parameter range of $b$, we consider the following set of dynamical equations:
\be
    3H^2=\rho_m+\frac{F}{4}+\frac{bF}{2\beta F +1}, \;\;\;\;\;\;\;\; -2\dot H=\rho_m+\frac{F}{3}\Big[1+\frac{4b}{(2F\beta+1)^2}\Big],\label{feq-c2}
\ee along with $\dot{F} = -4H F$ where we are considering ordinary matter $\rho_m$ (non-relativistic dust $\omega_m=0$).

To convert this set of second-order equations into a first-order dynamical system, we introduce the following variables to obtain an autonomous first-order system:
    
\be 
    x=\frac{F}{12H^2},\;\;\;\;\;\;\;\; y=\frac{H^2 \beta}{H^2 \beta+1},\label{newvars}
\ee which is bounded in such a way that the space phase is given by $\Psi=\{(x,y)\;\;\;|\;\;\;0\leq x\leq 1,\;\;\; 0\leq y\leq 1\}$.

The dynamical system obtained from Eqs. (\ref{feq-c2}) in the new variables (\ref{newvars}) takes the form:
\be
    x'=-2x(2+\frac{\dot H}{H^2}),\;\;\;\;\;\;\;\; y'=2y(1-y)\frac{\dot H}{H^2},\label{DynSys1}
\ee where
\be 
    \Omega_m=1-x-\frac{4bx(1-y)}{24xy+(1-y)}, \;\;\;\;\;\;\;\; -2\frac{\dot H}{H^2}=3\Omega_m+4x\Big[1+\frac{4b(1-y)^2}{(24xy+(1-y))^2}\Big] \label{Om1}.
\ee are the first and the second Friedmann (\ref{feq-c2}), respectively.

There are three critical points associated with this dynamical system:

\begin{enumerate}
    \item Matter-Dominated Critical Point $P_m(0,0)$ ($\Omega_m = 1$): At this point, the deceleration parameter is $q = 1/2$, and the effective barotropic parameter is $\omega_{eff} = 0$. The eigenvalues associated with this critical point are $\lambda_1 = -1$ and $\lambda_2 = -3$, indicating that this point is a decelerated future attractor. The squared sound speed is $c_s^2 = \frac{1}{3}$.

    \item Maxwell-Dominated Critical Point $P_M(1,1)$ ($\Omega_B = x =1$): This point is characterized by the deceleration parameter is $q = 0$, and the effective barotropic parameter is $\omega_{eff} = \frac{1}{3}$. The squared sound speed is $c_s^2 = \frac{1}{3}$, and the eigenvalues are $\lambda_1 = 1$ and $\lambda_2 = 4$, indicating that this point is a non-accelerated expanding past attractor. Since $y = 1$, the model starts with $H \to \infty$, confirming that it represents an early-time point.

    \item Maxwell/NLED Radiation Critical Point $P_{NL}\left(\frac{1}{4b+1}, 0\right)$ ($\Omega_M = \frac{1}{4b+1}$, and $\Omega_{NL} = \frac{4b}{4b+1}$): This point depends on the parameter $b$. The eigenvalues are $\lambda_1 = 1$ and $\lambda_2 = -4$, indicating a non-accelerated expanding saddle point where $\omega_{eff} = \frac{1}{3}$. When $b = 0$, this point reduces to the Maxwell-dominated point $P_M(1,0)$. This critical point remains within the phase space ($0 \leq x \leq 1$) only when $b > 0$. Considering the stability of the model, we only consider the parameter range $0 \leq b \leq \frac{108}{343}$.
\end{enumerate}

\begin{figure*}[h]
    \centering
    \includegraphics[width=8cm]{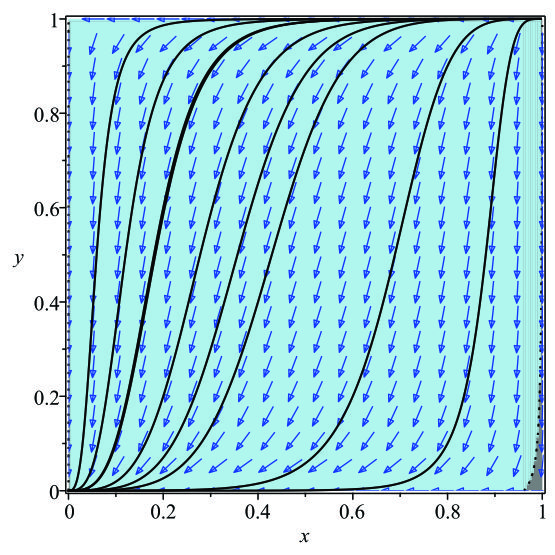}
    \includegraphics[width=8cm]{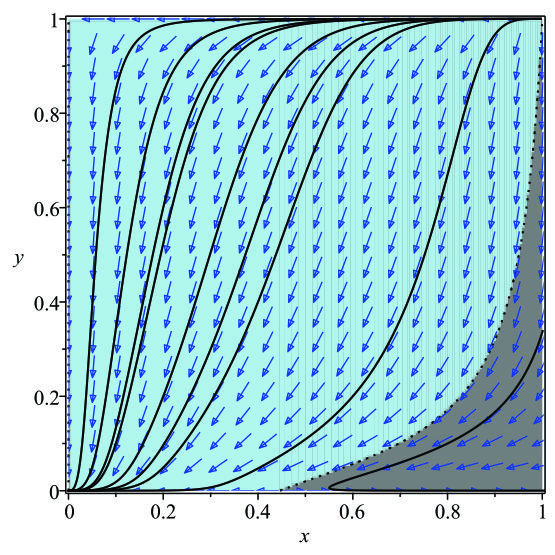}
    \caption{Phase portrait of the dynamical system (\ref{DynSys1}) for different parameter values $b$. The left panel shows $b=0.01$, and the right panel shows $b=0.31$. The gray region in the physical phase space (\ref{phsp-c1}) corresponds to the inequalities $0 \leq c^2_s \leq 1$, while the blue region corresponds to $0 \leq \Omega_m \leq 1$. These panels fulfill the conditions for stability.}
    \label{F51}
\end{figure*}

Figure \ref{F51} illustrates the entire phase space for two values of the parameter $b$ within the stability range. As observed, larger values of $b$ result in the critical saddle point being closer to zero.

\begin{figure*}[h]
    \centering
    \includegraphics[width=8cm]{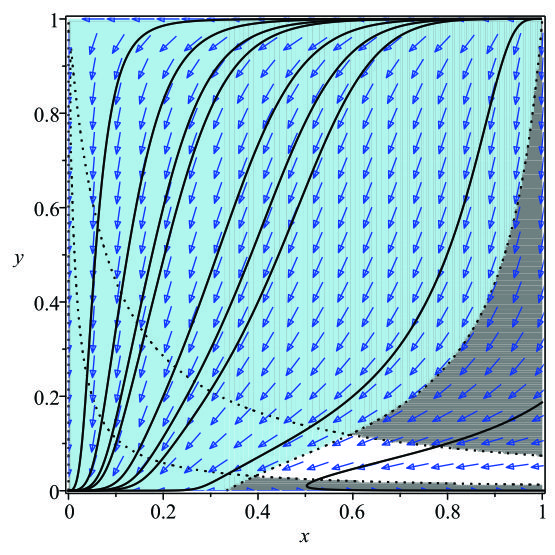}
    \caption{Phase portrait of the dynamical system (\ref{DynSys1}) for $b=0.5$ which  is outside the values for stability and causality. The gray region in the physical phase space (\ref{phsp-c1}) corresponds to the inequalities $0 \leq c^2_s \leq 1$, while the blue region corresponds to $0 \leq \Omega_m \leq 1$.}
    \label{F52}
\end{figure*}

In Figure \ref{F52}, the region of stability and causality is colored in gray. Hence, the model's evolution transitions from a stable and causal region and passes through a region between the two dotted lines (white region)—indicating a zone where it is neither stable nor causal—and re-enters a region where the model is viable.

The evolution of this model aligns with expectations: it begins in a Maxwell radiation-dominated domain, evolves towards a state dominated by non-relativistic matter, and transiently passes through a point dominated by non-linear radiation.

In conclusion, analyzing the rational nonlinear electrodynamics model with the Maxwell term given by the Lagrangian density (\ref{RNLED1}) has revealed a detailed picture of its dynamical behavior. We identified three critical points: a future attractor dominated by ordinary matter ($P_m$), a past attractor dominated by Maxwell radiation ($P_M$), and a saddle point influenced by both Maxwell and nonlinear electrodynamics ($P_{NL}$). The model demonstrates a consistent evolution starting from a Maxwell radiation-dominated phase, transitioning through a nonlinear radiation phase, and culminating in a state dominated by non-relativistic matter. 

None of the models we have analyzed can satisfactorily explain the current accelerated expansion of the universe. To account for the current stage of accelerated expansion, these models would require the inclusion of dark energy, either with some special material component or with modifications to the theory of relativity. Consequently, they fail to comprehensively explain this component of the cosmic background. This limitation diminishes their interest as independent models to describe the entire evolution of the universe, particularly in the context of dark energy.

\subsection{Generalized Rational Nonlinear Electrodynamics Dynamical Evolution with Maxwell Term}

In this section, we analyze the dynamical evolution of a model corresponding to the Lagrangian density given by (\ref{GRNLED}) with $\epsilon=1$. As previously discussed, constraining the values of the parameters of this Lagrangian density is highly complex.

The dynamical equations for this model are given by:
\bea
    3H^2&=&\rho_m+\frac{F}{4}+\frac{bF}{1+(2\beta F)^\alpha}, \nonumber \\
    -2\dot H&=&\rho_m-\frac{4F}{3}\Big[-\frac{1}{4}-\frac{b}{(1+ (2\beta F)^\alpha)}+\frac{b \alpha(2\beta F)^\alpha}{(1+ (2\beta F)^\alpha)^2}\Big], \label{feq-c2GN}
\eea where we are considering ordinary matter $\rho_m$ (non-relativistic dust $\omega_m=0$) and $\dot{F} = -4H F$.

We consider the following new variables:
\be 
    x=\frac{F}{3H^2},\;\;\;\;\;\;\;\; 
    y=\frac{(2\beta F)^\alpha}{1+(2\beta F)^\alpha}. \label{newvarsG}
\ee 
The phase space is defined by $\Psi = \{(x, y) \mid 0 \leq x \leq 1, \; 0 \leq y \leq 1\}$. Thus, the dynamical equations (\ref{feq-c2G}) in these new variables can be expressed as:
\begin{equation}
    x' = -2x(2+\frac{\dot{H}}{H^2}), \;\;\;\;\;
    y' = -4 \alpha y (1-y), \label{DynSysGRb}
\end{equation}
where
\bea
    \Omega_m&=&\frac{\rho_m}{3H^2}=1-x-4bx(1-y), \nonumber \\
    -2\frac{\dot H}{H^2}&=&3\Omega_m-16x\left[-\frac{1}{4}-b(1-y)+b \alpha y (1-y)\right]. \label{feq-c2G}
\eea

There are four critical points associated with this dynamical system:

\begin{enumerate}
    \item Matter-Dominated Critical Point $P_m(0,0)$ ($\Omega_m = 1$). At this point, $F \to 0$ implies negligible NLED effects, making this point representative of a purely matter-dominated phase. The deceleration parameter is $q = 1/2$, and the effective barotropic parameter is $\omega_{eff} = 0$. The eigenvalues associated with this critical point are $\lambda_1 = -1$ and $\lambda_2 = -4 \alpha$, indicating that this point is a decelerated future attractor if $\alpha > 0$ and saddle point if not. The squared sound speed is $c_s^2 = \frac{1}{3}$.

    \item Matter-Dominated Critical Point $P_m(0,1)$ ($\Omega_m = 1$) represents a matter-dominated phase with significant NLED effects, $F \to \infty$ . The deceleration parameter is $q = 1/2$, and the effective barotropic parameter is $\omega_{eff} = 0$. The eigenvalues associated with this critical point are $\lambda_1 = -1$ and $\lambda_2 = 4 \alpha$, indicating that this point is a saddle point if $\alpha > 0$ and decelerated future attractor if not. The squared sound speed is $c_s^2 = \frac{1}{3}$.

    \item Maxwell/NLED Radiation Critical Point $P_{NL}\left(\frac{1}{4b+1}, 0\right)$ ($\Omega_M = \frac{1}{4b+1}$, and $\Omega_{NL} = \frac{4b}{4b+1}$). This critical point remains within the phase space ($0 \leq x \leq 1$) only when $b > 0$ indicates that the NLED term is significant, whereas the Maxwell term may be less significant. The deceleration parameter is $q = 1$, and the effective barotropic parameter is $\omega_{eff} = 1/3$. The eigenvalues are $\lambda_1 = 1$ and $\lambda_2 = -4\alpha$, indicating a non-accelerated expanding saddle point if $\alpha > 0$ and past attractor if $\alpha < 0$. When $b=0$, this point reduces to the only Maxwell-dominated point $P_M(1,0)$, and when $b \to \infty$, it reduces to the matter-dominant point $P_m(0,0)$.

    \item Maxwell-Dominated Critical Point $P_M(1,1)$ ($\Omega_M =1$) due to the Maxwell term dominates the dynamics, implying that the effects of NLED are negligible or secondary. This point is characterized by the deceleration parameter being $q = 1$, and the effective barotropic parameter is $\omega_{eff} = 1/3$. The squared sound speed is $c_s^2 = \frac{1}{3}$, and the eigenvalues are $\lambda_1 = 1$ and $\lambda_2 = 4 \alpha$, indicating that this point is a non-accelerated expanding past attractor if $\alpha > 0$ and saddle point if not.
\end{enumerate}

\begin{figure*}[h]
    \centering
    \includegraphics[width=8cm]{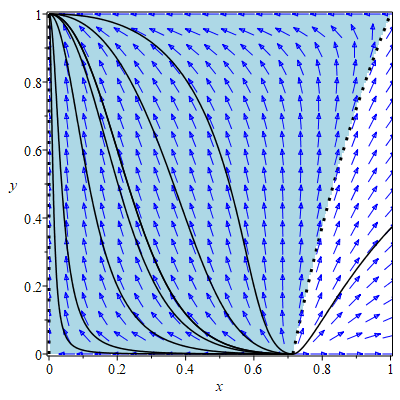}
    \includegraphics[width=8cm]{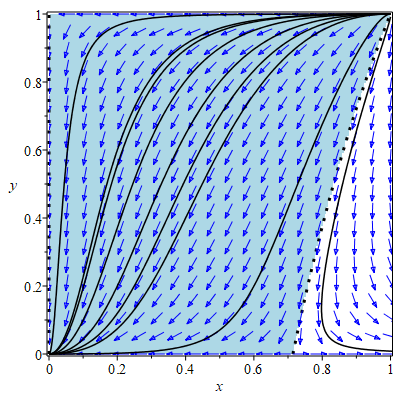}
    \caption{Phase portrait of the dynamical system (\ref{DynSys1}) for different parameter values $b$. The left panel shows $b=0.1$ and $\alpha=-1/2$. Whereas the right panel shows $b=0.1$ and $\alpha=1/2$. The blue region in the physical phase space corresponds to the inequalities $0 \leq \Omega_m \leq 1$. }
    \label{F6}
\end{figure*}

In the scenario where $\alpha < 0$, the phase space trajectories begin at the critical point $P_{NL}\left(\frac{1}{4b+1}, 0\right)$, which acts as a past attractor. This point represents an initial phase dominated by nonlinear electrodynamics (NLED) radiation, consistent with the hypothesis that nonlinear radiation could be more energetic than Maxwell radiation in the early universe. As the trajectories move towards the critical point $P_M(1,1)$, a saddle point, the nonlinearity of the electromagnetic field dilutes. Subsequently, the trajectories pass through the transitional critical point $P_{m1}(0,0)$, characterized by a dominance of ordinary matter. Finally, the trajectories converge to the critical point $P_{m2}(0,1)$, a future attractor dominated by ordinary matter with a certain electromagnetic component. This sequential transition from a phase dominated by nonlinear radiation to one dominated by Maxwell radiation, and ultimately to a matter-dominated state with an electromagnetic component, does not align well with the consensus cosmological model. This scenario reflects an orderly evolution of the universe through different energy domination phases, beginning with highly energetic NLED radiation, diluting through Maxwell radiation, and ending in a mixed matter and electromagnetic state. This scenario is shown in the left panel of Figure \ref{F6}.

In the scenario where $\alpha > 0$, the phase space trajectories begin at the critical point $P_M(1,1)$, which acts as a past attractor. This point represents an initial phase dominated by Maxwell radiation. The trajectories then move towards the critical point $P_{NL}\left(\frac{1}{4b+1}, 0\right)$, a saddle point where the dominance transitions to nonlinear electrodynamics (NLED) radiation. Subsequently, the trajectories pass through the transitional critical point $P_{m2}(0,1)$, characterized by ordinary matter dominance but with an essential component of Maxwell radiation. Finally, the trajectories converge to the critical point $P_{m1}(0,0)$, a future attractor dominated by ordinary matter. This pattern reflects a sequential transition from a Maxwell radiation-dominated phase to a nonlinear radiation-dominated phase, passing through a phase of pure ordinary matter and ultimately converging in a state dominated by ordinary matter with some electromagnetic components. This scenario is shown in the right panel of Figure \ref{F6}.

\section{Parameter Fitting with Observational Data}

Performing parameter fitting using observational data is essential to further validate and refine our models. Parameter fitting allows us to determine the values of the model parameters that best match the empirical data, providing a robust test of the theoretical models against real-world observations. This process is essential for several reasons. Firstly, it enables us to quantify the degree of agreement between the theoretical predictions and the observed data, thus assessing the viability of the models. Secondly, by constraining the parameter values, we can reduce the uncertainty in the model predictions, leading to more precise and reliable cosmological insights.

In this study, we use 31 cosmic chronometers collected over several years \cite{jimenez2003constraints, simon2005constraints, stern2010cosmic,  moresco2012new, zhang2014four, moresco2015raising, moresco20166, ratsimbazafy2017age}. Additionally, we use high-precision Baryon Acoustic Oscillation (BAO) measurements at different redshifts up to $z<2.36$ from BOSS DR14 quasars (eBOSS) \cite{ata2018clustering}, SDSS DR12 Galaxy Consensus \cite{alam2017clustering}, Ly-$\alpha$ DR14 cross-correlation \cite{blomqvist2019baryon}, Ly-$\alpha$ DR14 auto-correlation \cite{de2019baryon}, Six-Degree Field Galaxy Survey (6dFGS) \cite{beutler20116df}, and SDSS Main Galaxy Sample (MGS) \cite{ross2015clustering}. In addition, we include SNeIa data from the Pantheon compilation \cite{scolnic2018complete} with 1048 supernovae. We also incorporate a compressed version of Planck-15 information, treating the CMB as a BAO experiment at redshift $z=1090$, measuring the angular scale of the sound horizon \cite{aubourg2015}. These datasets provide a comprehensive and high-precision set of observations that span a significant range of the universe's history, making them ideal for testing cosmological models incorporating nonlinear electrodynamics (NLED). By fitting the parameters of our NLED models to these observations, we aim to derive constraints that will either support or challenge the theoretical framework, ultimately contributing to a deeper understanding of the universe's evolution.

\begin{figure*}[h!]
    \centering
    \makebox[12cm][c]{
    
    \includegraphics[width=8cm]{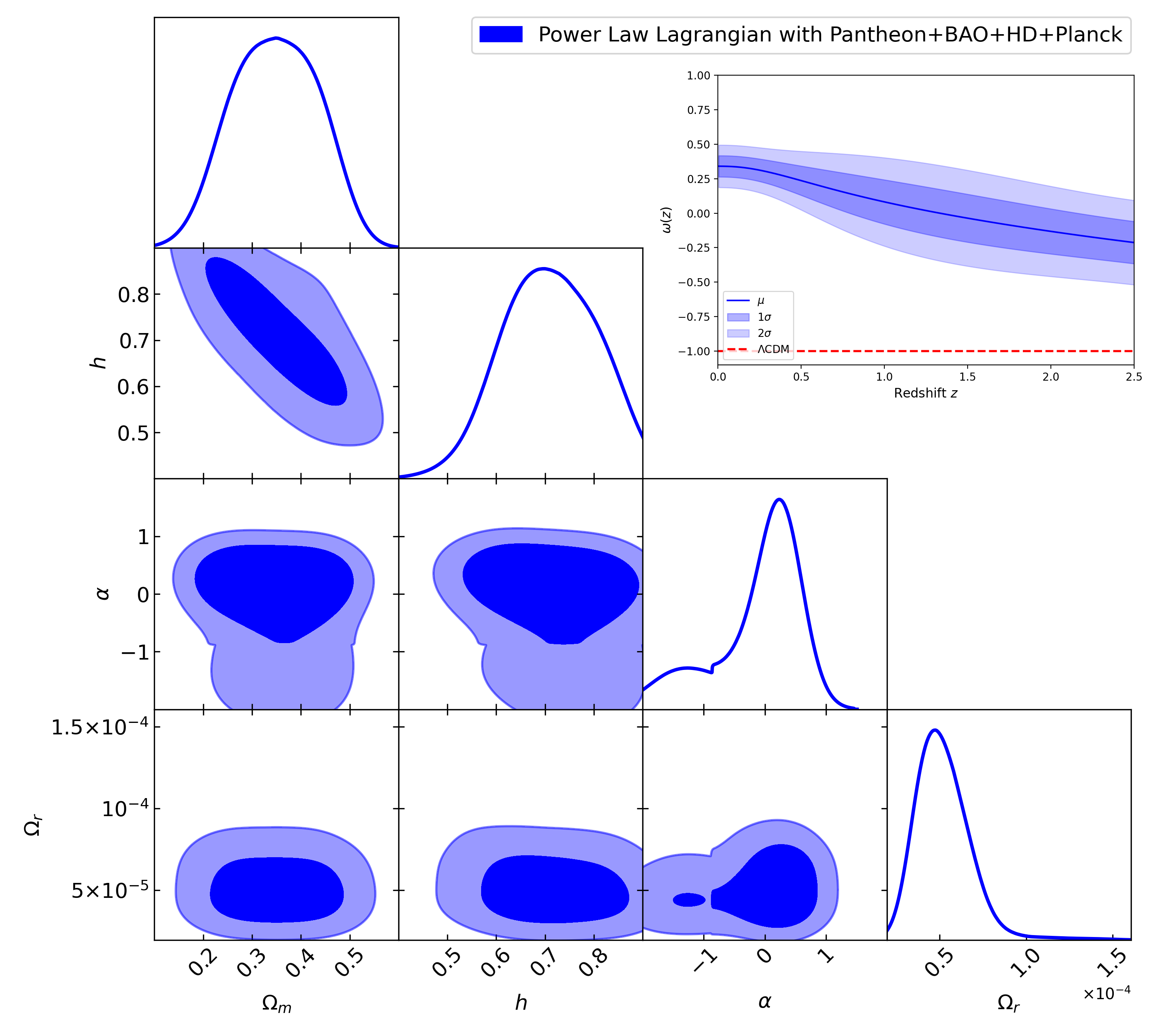}
        
    \includegraphics[width=8cm]{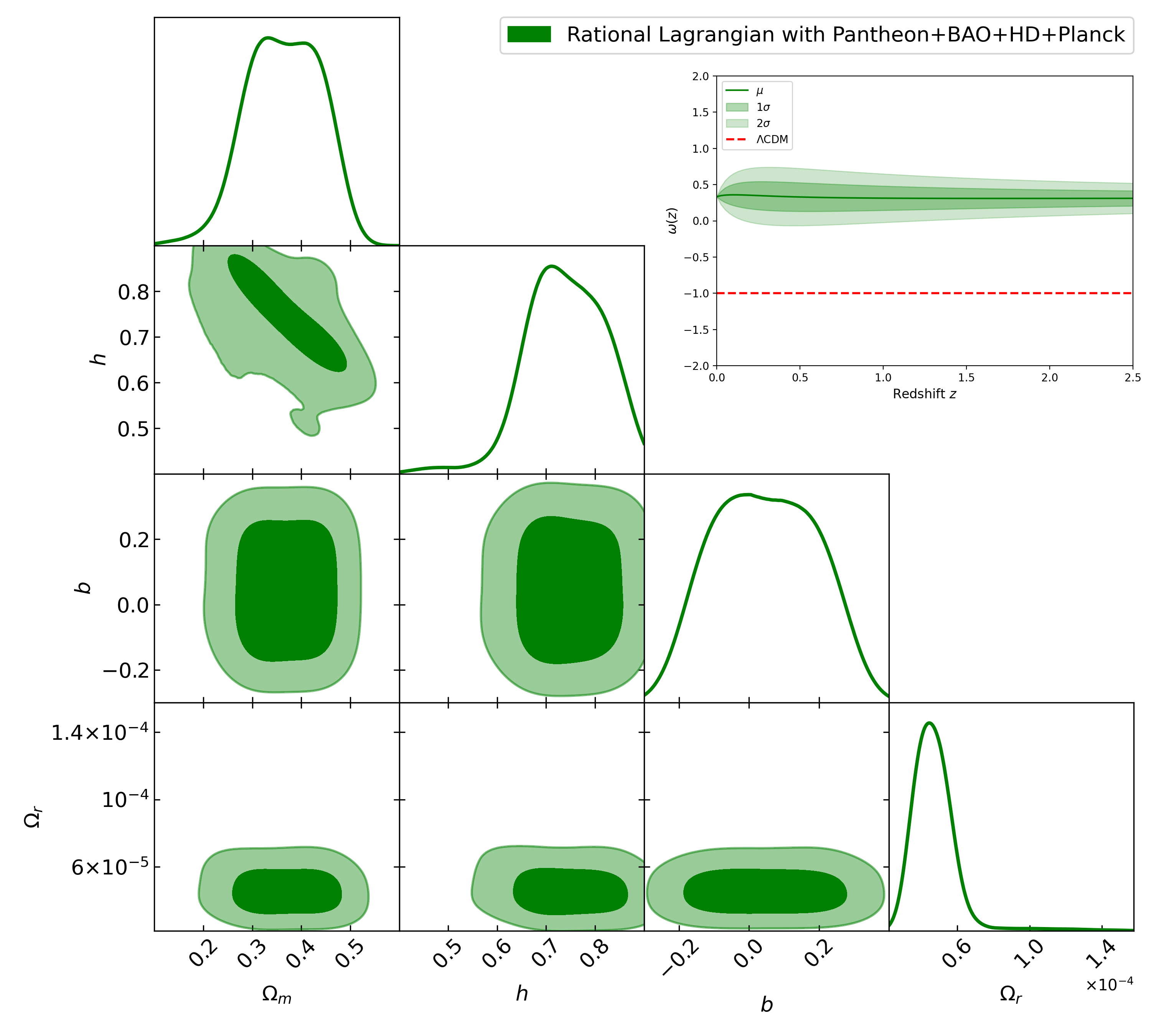}
    }
    \caption{Posterior sampling for the parameters using Pantheon, BAO, cosmic chronometers, and the Planck point for the models given in Eq. (\ref{Obsalpha})  and Eq. (\ref{Obsb}). The left panel shows the parameter estimation for the Power Law Lagrangian, while the right panel depicts the Rational Lagrangian model. The contour plots illustrate the constraints on $\Omega_m$, $h$, $\alpha$, $b$, and $\Omega_r$, with the 1$\sigma$ and 2$\sigma$ confidence regions shaded. The insets display the equation of state parameter $\omega(z)$ as a function of redshift $z$ for both models, indicating their deviation from the $\Lambda$CDM model at late times.}
    \label{fig:triangle_plots}
\end{figure*}

We employ Bayesian inference to determine the values of the model parameters that best fit the observational data. 
In this study, we utilize a nested sampling algorithm, using 500 live points, within the library \texttt{dynesty} \cite{speagle2020dynesty}, and the \texttt{SimpleMC} code \cite{simplemc}. The resulting posterior distributions provide constraints on the parameters, enabling us to assess the viability and precision of the NLED models in describing the universe's evolution. The results of the Bayesian inference are shown in Table \ref{tab:parameter_estimation} and in the plots of Figure \ref{fig:triangle_plots}. In Table \ref{tab:parameter_estimation}, as a reference, we include the parameter estimation of the $\Lambda$CDM model using the same data and parameters for nested sampling. This allows us, using the logarithm of the Bayesian evidence \(\log Z\), to calculate the Bayes factor \(\log B\), and through the Bayes factor and Jeffrey's scale \cite{Vazquez:2012ce, vazquez2020bayesian}, to make a model comparison.

Therefore, we employ Bayesian inference to investigate the parameters of the Power-Law and Rational Lagrangian models, aiming to constrain them using observational data from the late universe. Specifically, we aim to assess whether the nonlinear electrodynamics models are compatible with the available observational evidence and whether the results of our dynamical analysis align with the estimated parameters.

By fitting the parameters of the Power-Law and Rational Lagrangian models to the comprehensive observational datasets, we aim to derive constraints on the model parameters. This approach will allow us to evaluate the compatibility of the NLED models with current observational evidence and to determine whether these models can provide a viable description of the universe's evolution. In addition, we will quantify the degree of agreement between the theoretical predictions of the NLED models and the empirical data. The results of this analysis will provide critical insights into the viability and precision of the NLED models in the context of modern cosmology.

In the case of the Power-Law Lagrangian, the Friedmann equation (\ref{FFm1a}) useful for performing Bayesian inference is given by:
\be
    \frac{H^2}{H_0^2} = \Omega_{m0}(1+z)^{3(1+\omega_m)} + \Omega_{r0}(1+z)^4 + \Omega_{NL0}(1+z)^{4\alpha}, \label{Obsalpha}
\ee
where $\Omega_{m0} = \frac{\rho_{m0}}{3H_0^2}$, $\Omega_{r0} = \frac{F_0}{12H_0^2}$, and $\Omega_{NL0} = \gamma \frac{F_0^{\alpha}}{3H_0^2}$. We use the following flat priors: $\Omega_m \in [0.1, 0.5]$, $h \in [0.4, 0.9]$, and $\alpha \in [-2, 2]$.

For the Rational Lagrangian (\ref{RNLED1}), the equation to be used is:
\be
    \frac{H^2}{H_0^2} = \Omega_{m0}(1+z)^3 + \Omega_{r0}(1+z)^4 + 4b \frac{\Omega_{r0}(1+z)^4}{24 \beta H_0^2 \Omega_{r0}(1+z)^4 + 1}, \label{Obsb}
\ee
where $\Omega_{m0} = \frac{\rho_{m0}}{3H_0^2}$ and $\Omega_{r0} = \frac{F_0}{12H_0^2}$. We assume $\beta = 1$ and use the flat priors $\Omega_m \in [0.1, 0.5]$, $h \in [0.4, 0.9]$, and $b \in [-27/125, 108/343]$ for the Bayesian parameter estimation.


We observe that the datasets used cannot constrain the parameters $\alpha$ and $b$ effectively; however, the values of the matter density $\Omega_m$ and the Hubble parameter $h$ for both models are consistent with the expected values for the current Universe. The fact that the $\alpha$ and $b$ parameters can take any value within the prior ranges aligns with the results found in the dynamical analysis.

In the triangle plots of Figure \ref{fig:triangle_plots}, the posterior sampling for both Lagrangians and their respective $\omega(z)$ values, using equation (\ref{ParBar} for both cases, can be seen. According to the Bayes factor in Table \ref{tab:parameter_estimation}, the $\Lambda$CDM model is significantly preferred over the two NLED models when using late-universe observations. This is expected, as neither of the NLED models exhibits a cosmological constant behavior in the late Universe, as shown in the EoS plots in Figure \ref{fig:triangle_plots}.

\begin{table}[]
    \centering
    \resizebox{\textwidth}{!}{
    \begin{tabular}{|c |c |c | c |}
        \hline
                    &  Power Law          &   Rational & $\Lambda$CDM\\
         \hline
         \hline
         
         $\Omega_m$ & $0.3481 \pm 0.0866$ & $0.36637 \pm 0.0751$ & $0.3005 \pm 0.0064$\\
         \hline
         $h$       & $0.7133 \pm 0.09745$ &  $0.7413 \pm 0.0829$  & $0.6830 \pm 0.0049$\\
         \hline
         $\Omega_r$ & $5.16099 \times 10^{-5} \pm 1.5742 \times 10^{-5}$ & $4.7019 \times 10^{-5} \pm 1.3266 \times 10^{-5}$ & $5.3018 \times 10^{-5} \pm 7.6633 \times 10^{-7}$ \\
         \hline
         $\alpha$  & $-0.0679 \pm 0.6644$ &  -- & -- \\
         \hline
         $b$       & -- & $0.04566 \pm 0.1476$ & -- \\
         \hline
         $\log Z$  & $-1784.16 \pm 0.25$ & $-2355.57 \pm 0.22$ & $-538.37 \pm 0.215$ \\
         \hline
         \scriptsize{$\log B$ with $\Lambda$CDM} & $1244.8$ & $1817.2$ & -- \\
         \hline
         \scriptsize{Evidence over $\Lambda$CDM} & \scriptsize{Very strong in favour of $\Lambda$CDM} & \scriptsize{Very strong in favour of $\Lambda$CDM} & -- \\
         \hline
    \end{tabular}
    }
    \caption{Parameter estimation for the NLED Power-Law (Eq. \ref{Obsalpha}) and NLED Rational (Eq. \ref{Obsb}) Lagrangians. The table presents the estimated values of each model's matter density parameter $\Omega_m$, the Hubble parameter $h$, and the radiation density parameter $\Omega_r$. The parameter $\alpha$ is also estimated for the power-law model, while the parameter $b$ is included for the rational model. The $\log Z$ values indicate the Bayesian evidence for each model, and the $\log B$ values provide the Bayes factor comparison with the $\Lambda$CDM model, showing a strong preference for the $\Lambda$CDM model. The last row qualitatively describes the strength of evidence favoring $\Lambda$CDM over the NLED models.}

    \label{tab:parameter_estimation}
\end{table}

The Bayesian analysis results shown in Table \ref{tab:parameter_estimation} indicate that the $\Lambda$CDM model is favored significantly over the power-law and rational NLED models. The very strong evidence in favor of $\Lambda$CDM, indicated by the Bayes factor $\log B$, aligns with the expectations based on cosmological constant behavior in the late Universe, which neither NLED models can replicate. This suggests that while NLED models provide interesting theoretical insights, they may not be suitable for describing the current accelerated expansion of the Universe.

Moreover, the inability to tightly constrain the parameters $\alpha$ and $b$ within the prior ranges indicates that further theoretical refinement and possibly new observational data are needed to fully assess the viability of NLED models. The consistency of $\Omega_m$ and $h$ with expected values reinforces the reliability of the data and the robustness of the $\Lambda$CDM model as the standard cosmological model.

\section{Conclusions}

This study explored the dynamics of cosmological models incorporating nonlinear electrodynamics (NLED), focusing on their stability and causality. Employing a combination of dynamical systems theory and Bayesian inference, we analyzed two specific NLED models: the power law and the Rational Lagrangian models.

We identified stable and causal parameter ranges for the Power-Law Lagrangian model. In the dynamical systems analysis, this model transitions through various cosmological phases, from a Maxwell radiation-dominated state and evolving to a matter-dominated state. For the Rational Lagrangian model, including the Maxwell term, we observed stable and causal behavior for specific ranges of the parameter b. The phase space analysis revealed critical points indicating the evolutionary pathways of the universe, beginning from an early radiation-dominated state to an ordinary-matter-dominated state.

We performed Bayesian parameter estimation using a comprehensive set of observational data, including cosmic chronometers, Baryon Acoustic Oscillation (BAO) measurements, and Supernovae Type Ia (SNeIa). The estimated parameters for both models were consistent with the expected values for the current universe, particularly the matter density $\Omega_m$ and the Hubble parameter $h$. However, the parameters $\alpha$ and $b$ could not be tightly constrained within the prior ranges, which aligns with the findings from the dynamical analysis.

Based on Bayesian evidence, our model comparison strongly favored the $\Lambda$CDM model over the NLED models for late-universe observations. This result was anticipated, as neither NLED model exhibited an accelerated stage of cosmic expansion. None of the models we have analyzed can satisfactorily explain the current accelerated expansion of the universe. To account for the present stage of accelerated expansion, these models would require the inclusion of dark energy. Consequently, they fail to comprehensively explain this component of the cosmic background. This limitation diminishes their interest as standalone models for describing the entirety of the universe's evolution, particularly in the context of dark energy.

This investigation comprehensively explains the stability, causality, and dynamical evolution of cosmological models driven by nonlinear electrodynamics. While the NLED models exhibit intriguing theoretical properties, their compatibility with observational data suggests that further refinement and exploration are needed to fully integrate them into the standard cosmological framework. Future work should focus on extending the analysis to other forms of NLED and exploring their implications for early universe phenomena and high-energy astrophysical processes.

\section*{Acknowledgements}

RG-S acknowledges the support provided by SIP20230505-IPN and SIP20240638, FORDECYT-PRONACES-CONACYT CF-MG-2558591, COFAA-IPN, and EDI-IPN grants.

\bibliographystyle{unsrtnat}
\bibliography{PaperNLEDv1}

\end{document}